\documentclass[a4paper,11pt,
]{article}
\usepackage{amsmath,amssymb}
\usepackage[dvips]{color}
\usepackage{cite}
\usepackage{hangcaption}
\usepackage{amsmath,amssymb}
\usepackage[dvips]{color}
\usepackage{cite}
\usepackage{eepic}
\usepackage{epic}
\usepackage{hangcaption}
\usepackage{amsmath}
\usepackage{amssymb}
\usepackage{amscd}
\usepackage{amsthm}
\usepackage[dvipdfmx,%
 bookmarks=true,%
 bookmarksnumbered=true,%
 setpagesize=false,%
 pdfkeywords={TeX; dvipdfmx; hyperref; color;}]{hyperref}
\usepackage{hyperref}
\def\cases{\left\{\begin{array}{ll}}
\def\endcases{\end{array}\right.}

\def\bigtimes{\mathop{\mbox{\Large $\times$}}}
\begin{document}
\setcounter{page}{1}
\vskip1.5cm
\begin{center}
{\LARGE \bf
The double-slit quantum eraser experiments
and Hardy's paradox in the quantum linguistic interpretation
}
\vskip0.5cm
{\rm
\large
Shiro Ishikawa
}
\\
\vskip0.2cm
\rm
\it
\it
Department of Mathematics, Faculty of Science and Technology,
Keio University,
\\ 
3-14-1, Hiyoshi, Kouhoku-ku Yokohama, Japan.
\\
E-mail:
ishikawa@math.keio.ac.jp
\end{center}
\par
\rm
\vskip0.4cm
\par
\noindent
{\bf Abstract}
\normalsize
\par
\noindent 
Recently we proposed the linguistic quantum interpretation
(called quantum and classical measurement theory), which was characterized as a kind of metaphysical and linguistic turn of the Copenhagen interpretation.
This turn from physics to language does not only extend quantum theory to classical systems but also yield the quantum mechanical world view
(i.e., 
quantum philosophy or quantum language). 
The purpose of this paper
is to formulate
the double slits experiment,
the quantum eraser experiment,
Wheeler's delayed choice experiment,
Hardy's paradox
and the three boxes paradox
(the weak value associated with a weak measurement due to Aharonov, et al.)
in the linguistic interpretation of quantum mechanics.
Through these arguments, we assert that the linguistic interpretation is just 
the final version of so called Copenhagen interpretation.
And therefore, the Copenhagen interpretation does not belong to physics
(i.e.,
the realistic world view)
but the linguistic world view.

\par
\par
\vskip0.3cm
\par
\noindent
{\bf Keywords}: Copenhagen Interpretation, 
Many-worlds Interpretation,
Operator Algebra,
Quantum and Classical Measurement Theory,
Schr\"{o}dinger's Cat,
Wigner's Friend,
Double-Slit Experiment,
Quantum eraser experiment,
Wheeler's delayed choice experiment,
Hardy's paradox
and the three boxes paradox

%
\par
\vskip0.3cm
\par

\par
\def\Cal{\cal}
\def\bigstimes{\text{\large $\: \boxtimes \,$}}

\vskip1.0cm
\par
\noindent
\section{
Quantum language
(Axioms
and
Interpretation)
}
\subsection{The overview of quantum language}
\par
\noindent
\par
As mentioned in the above abstract, our purpose is to understand
the double slits experiment,
the quantum eraser experiment,
Wheeler's delayed choice experiment,
Hardy's paradox
and the three boxes paradox
in the linguistic interpretation of quantum mechanics, which is proposed in \cite{Ishi10}-\cite{Ishi22}.
\par
According to ref.\cite{Ishi14},
we shall mention the overview of quantum language
(or, measurement theory, in short, MT).
\rm
Quantum language is characterized as the linguistic turn of the Copenhagen interpretation of quantum mechanics.
Quantum language (or, measurement theory ) has two simple rules
(i.e. Axiom 1(concerning measurement) and Axiom 2(concerning causal relation))
and the linguistic interpretation (= how to use the Axioms 1 and 2). 
That is,
\begin{align}
\underset{\mbox{(=MT(measurement theory))}}{\fbox{Quantum language}}
=
\underset{\mbox{(measurement)}}{\fbox{Axiom 1}}
+
\underset{\mbox{(causality)}}{\fbox{Axiom 2}}
+
\underset{\mbox{(how to use Axioms)}}{\fbox{Linguistic interpretation}}
\label{eq1}
\end{align}
({\it cf.} refs.
{{{}}}{\cite{Ishi10}-\cite{Ishi22}}).
\par

%
%

\par
\par
\rm
Measurement theory
is,
by an analogy of
quantum mechanics
(or,
as a linguistic turn of quantum
mechanics
), constructed
as the scientific
theory
formulated
in a certain 
{}{$C^*$}-algebra ${\cal A}$
(i.e.,
a norm closed subalgebra
in the operator algebra $B(H)$
composed of all bounded linear operators on a Hilbert space $H$,
{\it cf.$\;$}\textcolor{black}{\cite{Neum, Saka}}
). Let ${\mathcal N}$ be the weak${}^\ast$ closure of ${\mathcal A}$,
which is called a $W^\ast$-algebra. The structure
$[{\mathcal A} \subseteq {\mathcal N} \subseteq B(H)]$
is called a fundamental structure of MT.
\par
\noindent
\par

When ${\cal A}={\mathcal C}(H)$,
the ${C^*}$-algebra composed
of all compact operators on a Hilbert space $H$,
the MT is called {quantum measurement theory}
(or,
quantum system theory),
which can be regarded as
the linguistic aspect of quantum mechanics.
Also, when ${\cal A}$ is commutative
$\big($
that is, 
when ${\cal A}$ is characterized by $C_0(\Omega)$,
the $C^*$-algebra composed of all continuous 
complex-valued functions vanishing at infinity
on a locally compact Hausdorff space $\Omega$
({\it cf.$\;$}\textcolor{black}{\cite{Saka}})$\big)$,
the MT is called {classical measurement theory}.
Thus, we have the following classification:
\begin{itemize}
\item[(A)]
$
\quad
\underset{\text{\scriptsize }}{\text{MT}}
$
$\left\{\begin{array}{ll}
\text{quantum MT$\quad$(when non-commutative ${\cal A}={\mathcal C}(H)$,
${\mathcal N}=B(H)$)}
%
\\
\\
\text{classical MT
$\quad$
(when commutative ${\cal A}=C_0(\Omega)$),
${\mathcal N}=L^\infty (\Omega, \nu)(\subseteq B(L^2(\Omega, \nu) )$)}
\end{array}\right.
$
\end{itemize}
%
%

\vskip0.7cm
\par
\noindent
Also, we assert that quantum language is located as follows:
\begin{center}
\begin{picture}(410,170)
{
\put(10,70){
{
\put(0,-3){
$\!\!
\underset{{\text{\footnotesize Alistotle}}}{\underset{{\text{\footnotesize Plato}}}{\overset{\text{\footnotesize Parmenides}}{{\overset{\text{\footnotesize Socrates}}{
{\fbox{\shortstack[l]{Greek\\ 
{\footnotesize philosophy}}}}
}
}
}
}
}
$
}
}
\put(51,-3){
\rm
$\xrightarrow[\text{\footnotesize sticism}]{\text{\footnotesize Schola-} }$
$\!\! \textcircled{\scriptsize 1}$
}
\put(93,7){
{\line(0,1){34}}
}
\put(93,-7){
{\line(0,-1){47}}
}
}
\put(100,70){
$
\begin{array}{l}
\!\!\!
{\; \xrightarrow[]{ \; 
\quad
}}
\overset{\text{\scriptsize (monism)}}{\underset{\text{\scriptsize (realism)}}
{\fbox{\text{Newton}}}}
{
\overset{\textcircled{\scriptsize 2}}{{\line(1,0){17}}}
}
\begin{array}{llll}
\!\!
\rightarrow
{\fbox{\shortstack[l]{relativity \\ theory}}}
{\xrightarrow[]{\qquad \quad \;\;\;}
}{\textcircled{\scriptsize 3}}
\\
\\
\!\!
\rightarrow
{\fbox{\shortstack[l]{quantum \\mechanics}}}
{
\xrightarrow[\qquad \quad \;\; ]{}
}\textcircled{\scriptsize 4}
\end{array}
\\
\\
\!\! \xrightarrow[]{{
\quad}}
\overset{\text{\scriptsize (dualism)}}{
\underset{\text{\scriptsize (idealism)}}{\fbox{
{\shortstack[l]{Descartes\\ Rock,... \\Kant}}
}}
}
{\xrightarrow[]{\textcircled{\scriptsize 6}
}}
\!
\overset{\text{\scriptsize (linguistic view)}}{\fbox{
\shortstack[l]{language \\ philosophy}
}}
\!\! \xrightarrow[{
}
{
\text{\footnotesize
zation
}}
]{{
{
}
{\text{\footnotesize
quanti-}
}
}}\textcircled{\scriptsize 8}
\end{array}
$
}
\put(300,86){
{\put(-40,0){\drawline(0,2)(0,-30)}}
{\put(-40,-32){\text{$\xrightarrow[]{\; \text{\footnotesize language}}$}}}
{\put(6,-32){\textcircled{\scriptsize 7}}}
}
\put(190,80){\line(0,1){46}}
\put(302,110){
$
\left.\begin{array}{llll}
\; 
\\
\; 
\\
\; 
\\
\;
\end{array}\right\}
\xrightarrow[]{\textcircled{\scriptsize 5}}
{\!\!\!
\overset{\text{\scriptsize (unsolved)}}{
\underset{\text{\scriptsize (quantum phys.)}}{
\fbox{\shortstack[l]{theory of \\ everything}}
}
}
}
$
}
\put(302,20){
$
\left.\begin{array}{lllll}
\; 
\\
\; 
\\
\; 
\\
\;
\\
\;
\end{array}\right\}
{\xrightarrow[]{\textcircled{\scriptsize 10}}}
\overset{\text{\scriptsize (=MT)}}{
\underset{\text{\scriptsize (language)}}{
\fbox{\shortstack[l]{\color{red}quantum\\ \color{red}language}}
}
}
$
}
\put(100,-70){
{\bf 
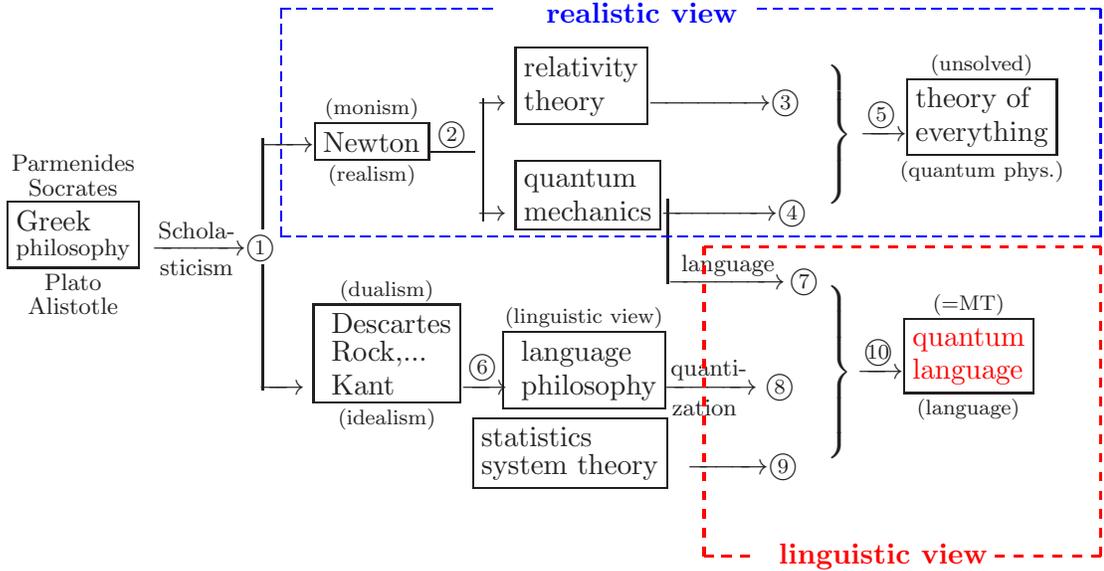
\hypertarget{Figure 1}{Figure 1}: The history of the world-view
}
}
\put(65,-32){
}
{
\thicklines
\color{red}
\dashline[50]{4}(287,-47)(270,-47)(270,70)(420,70)(420,-47)(380,-47)}
\thicklines
{\put(175,-16)
{{
{
\fbox{\shortstack[l]{ statistics \\ system theory}}
}
}$\xrightarrow[]{\qquad \;}$\textcircled{\scriptsize 9}
}
}
{
{
\put(288,-50){\color{red}
\bf
$\;\;$
linguistic view
}
\color{black}
}
{
\put(200,155){\color{blue}
\bf
$\;\;$
realistic view
}
\color{black}
}
}
}
{
\color{blue}
$\!\!\!\!\!\!\!\!${\dashline[50]{4}(190,160)(110,160)(110,74)(420,74)(420,160)(290,160)}
\color{black}
}
\end{picture}
\vskip2.8cm
\end{center}

\vskip1.0cm

\subsection{Observables}
\par
\noindent
\par
\par
\noindent
\par
Let
$[{\mathcal A} \subseteq {\mathcal N} \subseteq B(H)]$
be the fundamental structure
of measurement theory.
Let
${\cal N}_*$ be the
pre-dual Banach space of
${\cal N}$.
That is,
$ {\cal N}_* $
$ {=}  $
$ \{ \rho \; | \; \rho$
is a weak$^*$ continuous linear functional on ${\cal N}$
$\}$,
and
the norm $\| \rho \|_{ {\cal N}_* } $
is defined by
$ \sup \{ | \rho ({}F{}) |  \:{}: \; F \in {\cal N}
\text{ such that }\| F \|_{{\mathcal N}} 
(=\| F \|_{B(H)} )\le 1 \}$.
The bi-linear functional
$\rho(F)$
is
also denoted by
${}_{{\cal N}_*}
\langle \rho, F \rangle_{\cal N}$,
or in short
$
\langle \rho, F \rangle$.
Define the
\it
mixed state
$\rho \;(\in{\cal N}_*)$
\rm
such that
$\| \rho \|_{{\mathcal N}_* } =1$
and
$
\rho ({}F) \ge 0
\text{ 
for all }F\in {\cal N}
\text{ satisfying }
F \ge 0$.
And put
\begin{align*} {\frak S}^m  ({}{\cal N}_*{})
{=}
\{ \rho \in {\cal N}_*  \; | \;
\rho
\text{ is a mixed state}
\}.
\end{align*}
%
\rm

According to the noted idea ({\it cf.} ref. \textcolor{black}{\cite{ Davi}})
in quantum mechanics,
an {\it observable}
${\mathsf O}{\; \equiv}(X, {\cal F},$
$F)$ in the $W^*$-algebra
${{\cal N}}$
is defined as follows:
\par
\par
\begin{itemize}
\item[(B$_1$)]
[$\sigma$-field]
$X$ is a set,
${\cal F}
(\subseteq 2^X $,
the power set of $X$)
is a $\sigma$-field of $X$,
that is,
{\lq\lq}$\Xi_1, \Xi_2, \Xi_3, \cdots \in {\cal F}\Rightarrow \cup_{k=1}^\infty \Xi_k \in {\cal F}${\rq\rq},
{\lq\lq}$X \in {\mathcal F}${\rq\rq}
and
{\lq\lq}$\Xi  \in {\cal F}\Rightarrow X \setminus \Xi \in {\cal F}${\rq\rq}.
\item[(B$_2$)]
[Countably additivity]
$F$ is a mapping from ${\cal F}$ to ${{\cal N}}$ 
satisfying:
(a):
for every $\Xi \in {\cal F}$, $F(\Xi)$ is a non-negative element in 
${{\cal N}}$
such that $0 \le F(\Xi) $
$\le I$, 
(b):
$F(\emptyset) = 0$,
where
$0$ and $I$ is the $0$-element and the identity
in ${\cal N}$
respectively.
(c):
for any countable decomposition $\{ \Xi_1,\Xi_2, \ldots \}$
of $\Xi$
$\in {\cal F}$
(i.e., $\Xi_k , \Xi \in {\cal F}$
such that
$\bigcup_{k=1}^\infty \Xi_k = \Xi$,
$\Xi_i \cap \Xi_j= \emptyset
(i \not= j)$),
it holds that
\begin{align}
&
\quad
\lim_{K \to \infty } 
{}_{{}_{{\cal N}_*}}
\langle \rho, 
F( \bigcup_{k=1}^K \Xi_k )
 \rangle_{{}_{\cal N}}
=
{}_{{}_{{\cal N}_*}}
\langle \rho, 
 F( \Xi ) 
 \rangle_{{}_{\cal N}}
\quad
(
\forall \rho \in {\frak S}^m  ({\cal N}_*)
)
\label{eq2}
\end{align}
i.e.,
$
\lim_{K \to \infty }  
F( \bigcup_{k=1}^K \Xi_k )
= F( \Xi ) 
$
in the sense of weak${}^*$ convergence in ${\cal N}$.
\end{itemize}

\par
\noindent
\bf
Remark 1.
\rm
In the above (b), it is usual to assume the condition:
$F(X) = I$.
In fact, through all this paper except Section 5,
the condition:
$F(X) = I$ is assumed.
However,
for the reason mentioned in Remark 9 later, we start from the above (b).

%
%
\par
\vskip0.3cm
\par
\par
\noindent
%

\vskip1.0cm
\par
\subsection{Quantum language ( Axioms )}
\par
\noindent

\par
%
\par
\rm

With any {\it system} $S$, a fundamental structure
$[{\mathcal A} \subseteq {\mathcal N} \subseteq B(H)]$
can be associated in which the 
pure
measurement theory (A$_1$) of that system can be formulated.
A {\it pure state} of the system $S$ is represented by an element
$\rho^p (\in {\frak S}^p  ({}{\cal A}^*{})$="pure state class"({\it cf.} ref.\cite{Ishi14}))
and an {\it observable} is represented by an observable 
${\mathsf{O}}{\; =} (X, {\cal F}, F)$ in ${{\cal N}}$.
Also, the {\it measurement of the observable ${\mathsf{O}}$ for the system 
$S$ with the pure state $\rho^p$}
is denoted by 
${\mathsf{M}}_{{{\cal N}}} ({\mathsf{O}}, S_{[\rho^p]})$
$\big($
or more precisely,
${\mathsf{M}}_{\cal N} ({\mathsf{O}}{\; =} (X, {\cal F}, F), S_{[\rho^p]})$
$\big)$.
An observer can obtain a measured value $x $
($\in X$) by the measurement 
${\mathsf{M}}_{\cal N} ({\mathsf{O}}, S_{[\rho^p]})$.
\par
\noindent
\par
The Axiom$^{}$ 1 presented below is 
a kind of mathematical generalization of Born's probabilistic interpretation of quantum mechanics.
\par
\noindent
{\bf{Axiom$^{}$ 1\;\;
\rm
$[$Pure Measurement$]$}}.
\it
The probability that a measured value $x$
$( \in X)$ obtained by the measurement 
${\mathsf{M}}_{{{\cal N}}} ({\mathsf{O}}$
${ \equiv} (X, {\cal F}, F),$
{}{$ S_{[\rho^p_0]})$}
%
belongs to a set 
$\Xi (\in {\cal F})$ is given by
$
\rho^p_0( F(\Xi) )
$,
if $F(\Xi)$ is essentially continuous at $\rho^p_0$
({\it cf.} ref.\cite{Ishi14}).
\rm

\par
\par
\vskip0.2cm
\par

\par
Next, we explain Axiom 2 in (A$_1$).
Let $(T,\le)$ be a tree, i.e., a partial ordered 
set such that {\lq\lq$t_1 \le t_3$ and $t_2 \le t_3$\rq\rq} implies {\lq\lq$t_1 \le t_2$ or $t_2 \le t_1$\rq\rq}\!.
Assume that
there exists an element $t_0 \in T$,
called the {\it root} of $T$,
such that
$t_0 \le t$ ($\forall t \in T$) holds.
Put $T^2_\le = \{ (t_1,t_2) \in T^2{}\;|\; t_1 \le t_2 \}$.
The family
$\{ \Phi_{t_1,t_2}{}: $
${\cal N}_{t_2} \to {\cal N}_{t_1} \}_{(t_1,t_2) \in T^2_\le}$
is called a {\it causal relation}
({\it due to the Heisenberg picture}),
\rm
if it satisfies the following conditions {}{(C$_1$) and 
(C$_2$)}.
\begin{itemize}
\item[{\rm (C$_1$)}]
With each
$t \in T$,
a fundamental structure
$[{\mathcal A}_t \subseteq {\mathcal N}_t \subseteq B(H_t)]$
is associated.
\item[{\rm (C$_2$)}]
For every $(t_1,t_2) \in T_{\le}^2$, a continuous Markov operator 
$\Phi^{t_1,t_2}{}: {\cal N}_{t_2}{\mbox{(with the weak$^*$ topology)}}
$
$ \to
$
$ {\cal N}_{t_1}$
${\mbox{(with the weak$^*$ topology)}}$ 
is defined
(i.e.,
$\Phi^{t_1,t_2} \ge 0$,
$\Phi^{t_1,t_2}(I_{{\cal N}_{t_2}})$
$
=
$
$
I_{{\cal N}_{t_1}}$
).
And it satisfies that
$\Phi^{t_1,t_2} \Phi^{t_2,t_3} = \Phi^{t_1,t_3}$ 
holds for any $(t_1,t_2)$, $(t_2,t_3) \in T_\le^2$.
\end{itemize}
\noindent
The family of pre-dual operators
$\{ \Phi^{t_1,t_2}_*{}: $
$
{\frak S}^m  (({\cal N}_{t_1})_*)
\to {\frak S}^m  (({\cal N}_{t_2})_*)
\}_{(t_1,t_2) \in T^2_\le}$
is called a
{
\it
pre-dual causal relation}
({\it
due to the Schr\"{o}dinger picture}).
If we can regard that
$
\Phi^{t_1,t_2}_*{}
(
{\frak S}^p  (({\cal A}_{t_1})^*)
)
\subseteq
{\frak S}^p  (({\cal A}_{t_2})^*)
$,
the causal relation
$\{ \Phi_{t_1,t_2}{}: $
${\cal N}_{t_2} \to {\cal N}_{t_1} \}_{(t_1,t_2) \in T^2_\le}$
is said to be deterministic.

%

\par
\par
\rm
Now Axiom 2 in the measurement theory (\ref{eq1}) is presented
as follows:
\rm
\par
\noindent
{\bf{Axiom 2}
\rm[Causality]}.
\it
The causality is represented by
a causal relation 
$\{ \Phi^{t_1,t_2}{}: $
${\cal N}_{t_2} \to {\cal N}_{t_1} \}_{(t_1,t_2) \in T^2_\le}$.
\rm
\par
\par
\vskip0.2cm
\par
\noindent
\par

\par
\noindent
\par
\noindent
\par
\noindent
\vskip0.2cm
\par
\noindent
\subsection{
Linguistic Interpretation
}
\par
\noindent
\par
Next,
we have to
study the linguistic interpretation
(i.e.,
the manual of how to use the above axioms,
)
as follows.

That is, we present the following interpretation
(D)
[=(D$_1$), (D$_2$)],
which is characterized as a kind of linguistic turn
of so-called Copenhagen interpretation
({\it cf.} refs.\textcolor{black}{\cite{Ishi14, Ishi15}}
).
That is,
\begin{itemize}
\item[(D$_1$)]
Consider the dualism composed of {\lq\lq}observer{\rq\rq} and {\lq\lq}system( =measuring object){\rq\rq}.
And therefore,
{\lq\lq}observer{\rq\rq} and {\lq\lq}system{\rq\rq}
must be absolutely separated.
\item[(D$_2$)]
Only one measurement is permitted.
And thus,
the state after a measurement
is meaningless
$\;$
since it 
can not be measured any longer.
Therefore,
the wave collapse is prohibited.
Also, the causality should be assumed only in the side of system,
however,
a state never moves.
Thus,
the Heisenberg picture should be adopted.
And thus,
the Schr\"{o}dinger picture
is rather makeshift. Thus, the problem "when and where a measurement is performed?" is nonsense.
\end{itemize}
\par
\noindent
and so on.
For example, the axioms seem the rule of how to move the piece of a chess game.
On the other hand,
the linguistic interpretation resembles the standard tactics of chess
game.
In this sense,
we cannot completely say all about the linguistic interpretation.

%

\par
\noindent
\par
The following argument is a consequence of the above (D$_2$).
For each
$k=1,$
$2,\ldots,K$,
consider a measurement
${\mathsf{M}}_{{{\cal N}}} ({\mathsf{O}_k}$
${\; \equiv} (X_k, {\cal F}_k, F_k),$
$ S_{[\rho]})$.
However,
since
the (D$_2$)
says that
only one measurement is permitted,
the
measurements
$\{
{\mathsf{M}}_{{{\cal N}}} ({\mathsf{O}_k},S_{[\rho]})
\}_{k=1}^K$
should be reconsidered in what follows.
Under the commutativity condition such that
\begin{align}
&
F_i(\Xi_i) F_j(\Xi_j) 
=
F_j(\Xi_j) F_i(\Xi_i)
\label{eq3}
%
\\
&
\quad
(\forall \Xi_i \in {\cal F}_i,
\forall \Xi_j \in  {\cal F}_j , i \not= j),
\nonumber
\end{align}
we can
define the product observable
(or, simultaneous observable)
${\text{\large $\times$}}_{k=1}^K {\mathsf{O}_k}$
$=({\text{\large $\times$}}_{k=1}^K X_k ,$
$ \boxtimes_{k=1}^K {\cal F}_k,$
$ 
{\text{\large $\times$}}_{k=1}^K {F}_k)$
in ${\cal N}$ such that
\begin{align*}
({\text{\large $\times$}}_{k=1}^K {F}_k)({\text{\large $\times$}}_{k=1}^K {\Xi}_k )
=
F_1(\Xi_1) F_2(\Xi_2) \cdots F_K(\Xi_K)
\\
\;
(
\forall \Xi_k \in {\cal F}_k,
\forall k=1,\ldots,K
).
\qquad
\qquad
\nonumber
\end{align*}
Here,
$ \boxtimes_{k=1}^K {\cal F}_k$
is the smallest $\sigma$-field including
the family
$\{
{\text{\large $\times$}}_{k=1}^K \Xi_k
$
$:$
$\Xi_k \in {\cal F}_k \; k=1,2,\ldots, K \}$.
Then, 
the above
$\{
{\mathsf{M}}_{{{\cal N}}} ({\mathsf{O}_k},S_{[\rho]})
\}_{k=1}^K$
is,
under the commutativity condition (\ref{eq3}),
represented by the simultaneous measurement
${\mathsf{M}}_{{{{\cal N}}}} (
{\text{\large $\times$}}_{k=1}^K {\mathsf{O}_k}$,
$ S_{[\rho]})$.


\par
\par
Consider a finite tree
$(T{\; \equiv}\{t_0, t_1, \ldots, t_n \},$
$ \le )$
with the root $t_0$.
This is also characterized by
the map
$\pi: T \setminus \{t_0\} \to T$
such that
$\pi( t)= \max \{ s \in T \;|\; s < t \}$.
Let
$\{ \Phi^{t, t'} : {\cal N}_{t'}  \to {\cal N}_{t}  \}_{ (t,t')\in
T_\le^2}$
be a causal relation,
which is also represented by
$\{ \Phi_{\pi(t), t} : {\cal N}_{t}  \to {\cal N}_{\pi(t)}  \}_{ 
t \in T \setminus \{t_0\}}$.
Let an observable
${\mathsf O}_t{\; \equiv}
(X_t, {\cal F}_{t}, F_t)$ in the ${\cal N}_t$ 
be given for each $t \in T$.
Note that
$\Phi_{\pi(t), t}
{\mathsf O}_t$
$(
{\; \equiv}
(X_t, {\cal F}_{t},
\Phi_{\pi(t), t} F_t)$
)
is an observable in the ${\cal N}_{\pi(t)}$.
For the case that a tree $T$ is not finite,
see \cite{Ishi11}.

The pair
$[{\mathbb O}_T]
$
$=$
$[
\{{\mathsf O}_t \}_{t \in T}$,
$\{ \Phi^{t, t'} : {\cal N}_{t'}  \to {\cal N}_{t}  \}_{ (t,t')\in
T_\le^2}$
$]$
is called a
{\it sequential causal observable}.
%
%
For each $s \in T$,
put $T_s =\{ t \in T \;|\; t \ge s\}$.
And define the observable
${\widehat{\mathsf O}}_s
\equiv ({\text{\large $\times$}}_{t \in T_s}X_t, \boxtimes_{t \in T_s}{\cal F}_t, {\widehat{F}}_s)$
in ${\cal N}_s$
as follows:
\par
\noindent
\begin{align}
\widehat{\mathsf O}_s
&=
\left\{\begin{array}{ll}
{\mathsf O}_s
\quad
&
\!\!\!\!\!\!\!\!\!\!\!\!\!\!\!\!\!\!
\text{(if $s \in T \setminus \pi (T) \;${})}
\\
{\mathsf O}_s
{\text{\large $\times$}}
({}\bigtimes_{t \in \pi^{-1} ({}\{ s \}{})} \Phi_{ \pi(t), t} \widehat {\mathsf O}_t{})
\quad
&
\!\!\!\!\!\!
\text{(if $ s \in \pi (T) ${})}
\end{array}\right.
\label{eq4}
\end{align}
if
the commutativity condition holds
(i.e.,
if the product observable
${\mathsf O}_s
{\text{\large $\times$}}
({}\bigtimes_{t \in \pi^{-1} ({}\{ s \}{})} \Phi_{ \pi(t), t}
$
$\widehat {\mathsf O}_t{})$
exists)
for each $s \in \pi(T)$.
Using (\ref{eq4}) iteratively,
we can finally obtain the observable
$\widehat{\mathsf O}_{t_0}$
in ${\cal N}_{t_0}$.
The
$\widehat{\mathsf O}_{t_0}$
is called the realization
(or,
realized causal observable)
of
$[{\mathbb O}_T]$.
%

\noindent

%

\vskip0.5cm

\par
\noindent
\bf
Remark 2
\rm
[Particle or wave].
The argument about the 
"particle vs. wave"
is meaningless in quantum language.
As seen in the following table,
this argument is traditional:
\par
\noindent
\begin{center}
\begin{tabular}{|c||c|c|}
\hline
Theories $\setminus$ P or W & Particle(=symbol) & Wave(=
mathematical representation
) \\ \hline\hline
Aristotles & hyle & eidos \\ \hline
Newton mechanics & point mass & state (=(position, momentum)) \\ \hline
Statistics & population & parameter \\ \hline
Quantum mechanics & particle & state ($\approx$ wave function) \\ \hline
Quantum language & system (=measuring object) & state \\ \hline
\end{tabular}
\end{center}
In the above table,
Newtonian mechanics (i.e.,
mass point $\leftrightarrow$ state) may be easiest to understand.
Thus,
"particle" and "wave" are not confrontation concepts.
In this sense,
the
"wave or particle"
is meaningless.
In the linguistic interpretation of quantum mechanics, this should be usually understood as the problem
"interference or no interference".

\par
\noindent
\bf
Remark 3
\rm
[Reality].
Since quantum language is a kind of metaphysics,
we are not concerned with the reality such as 
discussed in \cite{Eins} and \cite{Bell}.
Also, since space and time are independent in quantum language
({\it cf.}
\cite{Ishi15}
),
we can not expect it to yield a good physical theory
(i.e., \textcircled{\footnotesize 5} in \hyperlink{Figure 1}{Figure 1}).

\par
\noindent
\bf
Remark 4
\rm
[The Schr\"{o}dinger's cat].
Axiom 2 allows us to deal with more than 
the deterministic causal relation,
for example,
the Brownian motion and the quantum decoherence,
etc.
Therefore,
we can easily describe the Schr\"{o}dinger's cat
by quantum language.
Thus, this is not a paradox in quantum language.
However, quantum language (due to dualism composed of {\lq\lq}observer{\rq\rq} and {\lq\lq}system{\rq\rq}) does not have a power to describe Wigner's friend as well as Descartes' proposition "I think, therefore I am"
({\it cf}. \cite{Ishi16}).
\par
\par
\noindent
\bf
Remark 5
\rm
[The commutative condition (\ref{eq3})].(i):
If the commutative condition (\ref{eq3})
does not hold,
${\text{\large $\times$}}_{k=1}^K {\mathsf{O}_k}$
$=({\text{\large $\times$}}_{k=1}^K X_k ,$
$ \boxtimes_{k=1}^K {\cal F}_k,$
$ 
{\text{\large $\times$}}_{k=1}^K {F}_k)$
is not an observable,
but
the ${\mathcal N}$-valued measure space.
\par
\noindent
(ii): Assume that there exists an observable
${\widehat{\mathsf{O}}}$
$=({\text{\large $\times$}}_{k=1}^K X_k ,$
$ \boxtimes_{k=1}^K {\cal F}_k, \widehat{F})$
such that
$$
\widehat{F}(X_1 \times X_2 \times \cdots \times X_{j-1} \times \Xi_j \times X_{j+1} \times \cdots X_K )=F_j(\Xi_j)
\quad
(\Xi_j \in {\mathcal F}_j, \;\; j=1,2,...,K)
$$
Then, there is a reason to call the ${\widehat{\mathsf{O}}}$ a 
\it
simultaneous observable.
\rm
\par
\noindent
(iii): Also, it may be worth while investigating the concept such that
${\widehat{\mathsf{O}}}$
$=({\text{\large $\times$}}_{k=1}^K X_k ,$
$ \boxtimes_{k=1}^K {\cal F}_k, \widehat{F})$
is 
\it
an simultaneous observable concerning $\rho$.
\rm

\par

\section{The double-slit experiment}
Although Feynmann's enthusiasm is transmitted in the explanation of
the double-slit experiment 
in \cite{Feyn},
we do not think that
his explanation is sufficient.
That is because 
the double-slit experiment and so on
should be explained after the answer to
"What kind of measurement is taken?".

\unitlength=0.7mm
\begin{picture}(150,110)
\put(0,50){\vector(1,0){180}}
\put(31,10){\vector(0,1){90}}
\put(180,50){$\;\; x$}
\put(24,100){$\;\; y$}
\put(25,49){P$\; \underset{\rightarrow}{\bullet}$ }
\allinethickness{0.7mm}
\put(15,50){\line(5,2){50}}
\put(15,50){\line(5,-2){50}}
\put(37,50){\line(5,2){28}}
\put(37,50){\line(5,-2){28}}
\put(65,70){\line(0,1){30}}
\put(65,50){\line(0,1){11}}
\put(65,50){\line(0,-1){11}}
\put(65,30){\line(0,-1){20}}
\put(67,47){$a$}
\put(67,66){$A$}
\put(67,32){$B$}
\multiput(120,50)(0,7){7}{\line(1,0){55}}
\dottedline{3}(140,97)(140,105)
\dottedline{3}(140,10)(140,15)
\multiput(120,50)(0,-7){5}{\line(1,0){55}}
\put(113,39){$(b,- \delta)$}
\put(116,46){$b$}
\put(113,53){$(b,\delta)$}
\put(113,60){$(b, 2\delta)$}
\end{picture}
\begin{center}
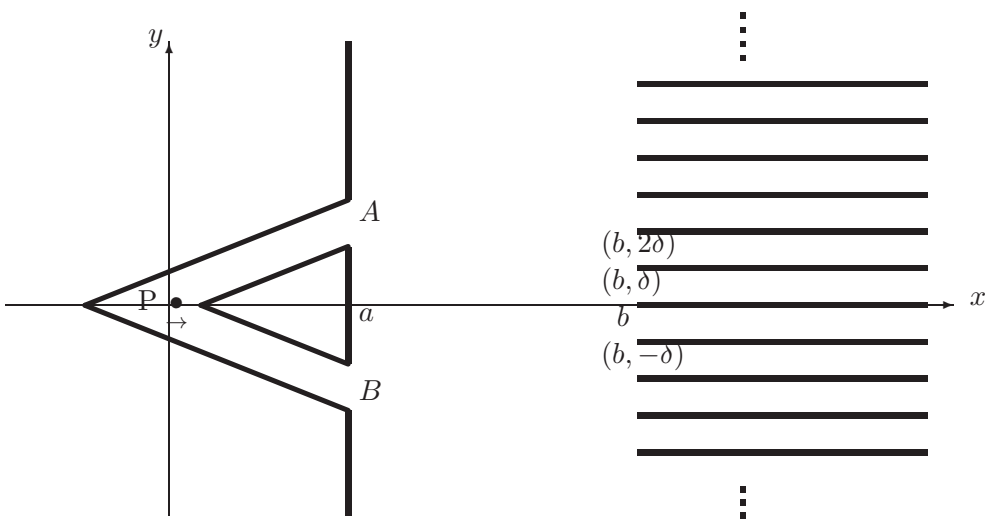
Figure 2(1). Potential $V_1(x,y)=\infty$ on the thick line,
$=0$ (elsewhere) 
\end{center}

\unitlength=0.7mm
\begin{picture}(150,110)
\put(0,50){\vector(1,0){180}}
\put(31,10){\vector(0,1){90}}
\put(180,50){$\;\; x$}
\put(24,100){$\;\; y$}
\put(25,49){P$\; \underset{\rightarrow}{\bullet}$ }
\allinethickness{0.7mm}
\put(15,50){\line(5,2){50}}
\put(15,50){\line(5,-2){50}}
\put(37,50){\line(5,2){28}}
\put(37,50){\line(5,-2){28}}
\put(65,70){\line(0,1){30}}
\put(65,50){\line(0,1){11}}
\put(65,50){\line(0,-1){11}}
\put(65,30){\line(0,-1){20}}
\put(65,50){\line(1,0){60}}
\put(67,47){$a$}
\put(67,66){$A$}
\put(67,32){$B$}
\multiput(120,50)(0,7){7}{\line(1,0){55}}
\dottedline{3}(140,97)(140,105)
\dottedline{3}(140,10)(140,15)
\multiput(120,50)(0,-7){5}{\line(1,0){55}}
\put(113,39){$(b,- \delta)$}
\put(116,46){$b$}
\put(113,53){$(b,\delta)$}
\put(113,60){$(b, 2\delta)$}
\end{picture}
\begin{center}
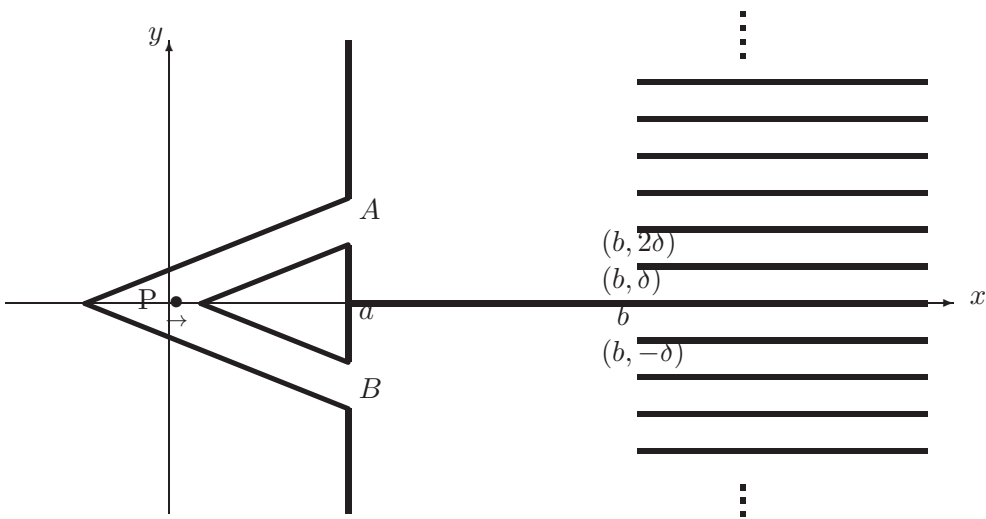
Figure 2(2). Potential $V_2(x,y)=\infty$ on the thick line,
$=0$ (elsewhere) 
\end{center}
That is, 
$$
V_2 = V_1 + \mbox{"the line segment $\overline{ab}$"}
$$

Consider a tree $(T, \le )$ with the two branches such that
$$
T=\{0\} \cup T_1 \cup T_2
$$
where
\begin{align*}
&
T_1=\{(1,s) \;|\; s > 0 \},
\qquad
T_2= \{(2,s) \;|\; s > 0 \}
\\
&
0 \le (i ,s_i) \qquad ( i=1,2, \;\; 0< s_i  )
\\
&
(i,s_i) \le (i,s'_i ) \qquad (i=1,2, \;\; s_i \le s'_i)
\end{align*}
\par
For each $t \in T$, define the fundamental structure
$$
[{\mathcal C}(H_t)
\subseteq B(H_t) \subseteq B(H_t)]
$$
where $H_t=L^2({\mathbb R}^2 )$
$(\forall t \in T)$.
Let $u_0 \in H_0 =  L^2({\mathbb R}^2 )$ be a initial wave-function
such that
($k_0>0$, small $\sigma > 0$):
$$
u_0(x,y)
\approx
\psi_x(x,0)
\psi_y (y,0)
=
\frac{1}{\sqrt{\pi^{1/2}\sigma }}
\exp \Big(ik_0 x -\frac{x^2}{2\sigma ^2}\Big)
\cdot
\frac{1}{\sqrt{\pi^{1/2}\sigma }} \exp \Big(-\frac{y^2}{2\sigma^2}\Big)
$$
where
the average momentum $(p^0_1,p^0_2)$ is calculated by
$$
(p^0_1,p^0_2)=
\Big(
\int_{\mathbb R} {\overline \psi}_x(x,0)
\cdot 
\frac{\hbar \partial \psi_x(x,0)}{i \partial x}
dx,
\int_{\mathbb R} {\overline \psi}_y(y,0)
\cdot 
\frac{\hbar \partial \psi_y(y,0)}{i \partial y}
dy
\Big)
=
(\hbar k_0, 0 )
$$
That is, we assume that the initial state
of the particle $P$
(
in
Figures 2(1) and 2(2)
)
is equal to
$| u_0 \rangle \langle u_0|$.

As mentioned in the above,
consider two branches
$T_1$ and $T_2$.
Thus, concerning $T_1$, we have the following
Schr\"{o}dinger equation:
$$i \hbar \frac{\partial}{\partial t }
\psi_t(x,y)
=
{\mathcal H}_1 \psi_t(x,y),
\qquad
{\mathcal H}_1
=
-\frac{\hbar^2}{2m}\frac{\partial^2}{\partial x^2}
-
\frac{\hbar^2}{2m}\frac{\partial^2}{\partial y^2}
+ V_1(x,y)
$$
Also, concerning $T_2$, we have the following
Schr\"{o}dinger equation:
$$i \hbar \frac{\partial}{\partial t }
\psi_t(x,y)
=
{\mathcal H}_2 \psi_t(x,y),
\qquad
{\mathcal H}_2
=
-\frac{\hbar^2}{2m}\frac{\partial^2}{\partial x^2}
-
\frac{\hbar^2}{2m}\frac{\partial^2}{\partial y^2}
+ V_2(x,y)
$$

Let $s_1, s_2 $ be sufficiently large positive numbers.
Put
$t_1=(1,s_1) \in T_1$,
$t_2=(2,s_2) \in T_2$.
Define the subtree $T' (\subseteq T)$ such that
$T'=\{0, t_1, t_2 \}$
and $0 < t_1$, $0 < t_2$.
Thus, we have the causal relation:
$
\{
\Phi_1^{0, t_i}: B(H_{s_i}) \to B(H_0)
\}_{i=1,2}
$
where
\begin{align*}
&
\Phi_1^{0, t_1} F =e^{
\frac{{\mathcal H}_1 s_1 }{i \hbar }
}
F_1
e^{-
\frac{{\mathcal H}_1 s_1}{i \hbar }
}
\qquad
(\forall F_1 \in B(H_{t_1})=B(L^2({\mathbb R}^2 )  )
)
\\
&
\Phi_2^{0, t_2} F =e^{
\frac{{\mathcal H}_2 s_2}{i \hbar }
}
F_1
e^{-
\frac{{\mathcal H}_2 s_2}{i \hbar }
}
\qquad
(\forall F_2 \in B(H_{t_2} )=B(L^2({\mathbb R}^2 )
))
\end{align*}

\par
Put ${\mathbb Z}$
$=$
$\{0, \pm 1, \pm 2, \cdots \}$.
Let $\delta$ be a sufficiently small positive number.
For each
$n \in {\mathbb Z}$,
define the region $D_n (\subseteq {\mathbb R}^2 )$
such that
\begin{align*}
D_0
&
=
\{ (x,y) \in {\mathbb R}^2 \;|\;
x <b \}
\\
\\
D_n
&
=
\begin{cases}
\{ (x,y) \in {\mathbb R}^2 \;|\;
b \le x , \; \delta(  n-1) < y \le \delta n  \}
\qquad &
(n=1,2, \cdots )
\\
\\
\{ (x,y) \in {\mathbb R}^2 \;|\;
b \le x ,\;  \delta n < y \le \delta( n+1)  \}
\qquad 
&
(n=-1,-2, \cdots )
\end{cases}
\end{align*}
Define the observable $({\mathbb Z}, 2^{{\mathbb Z}}, F )$
in $B(L^2({\mathbb R}^2)$
such that
$$
[F(\{n\})](x,y)
=
\chi_{_{D_n}} (x,y )
\quad
(\forall n \in {\mathbb Z}, \forall (x,y) \in {\mathbb R}^2 )
$$
where
$\chi_{_{D_n}} (x,y )=1 \;\; ((x,y)\in D_n ), \;=0 $
(elsewhere).

Hence,
we can consider the two observables
${\mathsf O}_{t_1}=({\mathbb Z}, 2^{{\mathbb Z}}, F )$
in $B(H_{t_1})
(=
B(L^2({\mathbb R}^2)
)$
and
${\mathsf O}_{t_2}=({\mathbb Z}, 2^{{\mathbb Z}}, F )$
in $B(H_{t_2})(=
B(L^2({\mathbb R}^2)
)$.

Since
$\Phi_1^{0,t_1} {\mathsf O}_{t_1}
=
({\mathbb Z}, 2^{{\mathbb Z}}, \Phi_1^{0,t_1}F )$
is the observable in $B(H_0)$,
we have the measurement
\begin{align}
{\mathsf M}_{B(H_0)}(\Phi_1^{0,t_1} {\mathsf O}_{t_1}, S_{[\rho_0]})
\label{eq5}
\end{align}
We consider that this is just the description of the standard double-slit experiment.
The following is well known:
\begin{itemize}
\item[(E$_1$)]
The measured date
$(x_1, x_2. \cdots, x_K) \in {\mathbb Z}^k$
obtained
by the parallel measurement
$\otimes_{k=1}^K {\mathsf M}_{B(H_0)}$
$(\Phi_1^{0,t_1} {\mathsf O}_{t_1}, S_{[\rho_0]})$
({\it cf}. \cite{Ishi14})
will show the interference fringes.
\end{itemize}
Also, since
$\Phi_2^{0,t_2} {\mathsf O}_{t_2}
=
({\mathbb Z}, 2^{{\mathbb Z}}, \Phi_2^{0,t_2}F )$
is the observable in $B(H_0)$,
we have the measurement
\begin{align}
{\mathsf M}_{B(H_0)}(\Phi_2^{0,t_2} {\mathsf O}_{t_2}, S_{[\rho_0]})
\label{eq6}
\end{align}
Fig. 2(2) says that
\begin{itemize}
\item[(E$_2$)]
if we get the positive measured value $n$ by the measurement
${\mathsf M}_{B(H_0)}(\Phi_2^{0,t_2} {\mathsf O}_{t_2}, S_{[\rho_0]})$,
we may conclude that
the particle $P$ passed through the hole $A$.
\end{itemize}
Further, note that we have the sequential causal observable
$[{\mathbb O}_{T'}]
$
$=$
$[
\{{\mathsf O}_{t_i} \}_{i=1,2}$,
$\{ \Phi_i^{0, t_i} : B(H_{t_i})  \to B(H_0)  \}_{i=1,2}
]$.
However, it should be noted that
\begin{itemize}
\item[(E$_3$)]
the sequential causal observable
$[{\mathbb O}_{T'}]
$
can not be realized,
since
the commutativity does not generally hold,
that is,
it generally holds that
$$
\Phi_1^{0, t_1}F(\Xi)
\cdot
\Phi_2^{0, t_2}F(\Gamma)
\not=
\Phi_2^{0, t_2}F(\Gamma)
\cdot
\Phi_1^{0, t_1}F(\Xi)
\qquad
(\forall \Xi, \Gamma \in 2^{\mathbb Z} )
$$
\end{itemize}

\bf
\par
\noindent
Remark 6
\rm
Although, strictly speaking, we have to say
that
the statement
"the particle $P$ passed through the hole $A$"
can not be described in terms of quantum language,
it should be allowed to say the statement (E$_2$).
Also, concerning the statement
(E$_3$),
note that
$$
{\mathsf O}_{t_1}=({\mathbb Z}, 2^{{\mathbb Z}}, F )={\mathsf O}_{t_2},
$$
but
the observables ${\mathsf O}_{t_1}$ and ${\mathsf O}_{t_2}$ are in different worlds
(i.e.,
different branches),
except while $\Phi_1^{0, t_1}=
\Phi_2^{0, t_2}$.
We consider that,
the double-slit experiment can not be completely explained without branches
In this sense,
our argument may be similar to Everett's
({\it cf.}
\cite{Ever}).
Also, for our other understanding of the double-slit experiment,
see \cite{Ishi8} and \cite{Ishi9}.

\section{The quantum eraser experiment}
\subsection{Usual situation}

Let $H$ be a Hilbert space.
And let ${\mathsf O}=(X, {\mathcal F}, F )$
be an observable in $B(H)$.
Let
$u_1$
and $u_2$
$( \in H)$ be orthonormal elements,
i.e.,
$\| u_1 \|_H=\| u_2 \|_H=1$
and
$\langle u_1 , u_2 \rangle =0$.
Put
$$
u=
\alpha_1 u_1 + \alpha_2  u_2
$$
where $\alpha_i \in {\mathbb C}$
such that
$| \alpha_1|^2 + |\alpha_2|^2=1$.

Consider the measurement:
\begin{align}
{\mathsf M}_{B( H )}
(
{\mathsf O}  , S_{[| u \rangle \langle u |]} ).
\label{eq7}
\end{align}
Then, the probability that a measured value
$ x(\in X )$
belongs to
$
\Xi (\in {\mathcal F} )$
is given by
\begin{align*}
&
\langle
u,
F(\Xi) 
u
\rangle
\\
=
&
\langle
\alpha_1  u_1  + \alpha_2  u_2,
 F(\Xi) 
(
\alpha_1  u_1  + \alpha_2 u_2
)
\rangle
\\
=
&
|\alpha_1|^2
\langle
u_1 ,
F(\Xi)u_1
\rangle
+
|\alpha_2|^2
\langle
u_2 ,
F(\Xi)u_2
\rangle
+
\overline{\alpha}_1\alpha_2
\langle
u_1 ,
F(\Xi)u_2
\rangle
+
{\alpha}_1 \overline{\alpha}_2
\langle
u_2 ,
F(\Xi)u_1
\rangle
\\
=
&
|\alpha_1|^2
\langle
u_1 ,
F(\Xi)u_1
\rangle
+
|\alpha_2|^2
\langle
u_2 ,
F(\Xi)u_2
\rangle
+
2
\mbox{[Real part]}
(
\overline{\alpha}_1\alpha_2
\langle
u_1 ,
F(\Xi)u_2
\rangle
)
\end{align*}
where
the
interference term
(i.e.,
the third term)
appears.

\subsection{Tensor Hilbert space}

Let ${\mathbb C}^2$ be the two dimensional Hilbert space,
i,e.,
${\mathbb C}^2$
$=$
$
\Big\{ 
\bmatrix
z_1
\\
z_2
\\
\endbmatrix
\;|
\;
z_1, z_2 \in {\mathbb C}
\Big\}$.
And put
$$
e_1 =\bmatrix
1
\\
0
\\
\endbmatrix,
\qquad
e_2 =\bmatrix
0
\\
1
\\
\endbmatrix
$$
Here,
define the observable
${\mathsf O}_x=(\{-1, 1 \}, 2^{\{-1, 1 \}}, F_x )$
in
$B({\mathbb C}^2)$
such that
$$
F_x(\{ 1 \} ) = \frac{1}{2}
\bmatrix
1 & 1
\\
1 & 1
\\
\endbmatrix,
\quad
F_x(\{ -1 \} ) = \frac{1}{2}
\bmatrix
1 & -1
\\
-1 & 1
\\
\endbmatrix,
$$
Here, note that
\begin{align*}
&
F_x(\{ 1 \} )e_1 = \frac{1}{2}(e_1 + e_2 ),
\quad
F_x(\{ 1 \} )e_2 = \frac{1}{2}(e_1 + e_2 )
\\
&
F_x(\{ -1 \} )e_1 = \frac{1}{2}(e_1 - e_2 ),
\quad
F_x(\{ -1 \} )e_2 = \frac{1}{2}(-e_1 + e_2 )
\end{align*}

Also, define the {\it existence} observable
${\mathsf O}_E=(\{ 1 \}, 2^{\{ 1 \}}, F_E )$
in
$B({\mathbb C}^2)$
such that
$$
F_E(\{ 1 \} ) = 
\bmatrix
1 & 0
\\
0 & 1
\\
\endbmatrix
%
$$
Further, define $\psi \in {\mathbb C}^2 \otimes H  $
$($
the tensor Hilbert space of ${\mathbb C}^2$ and $H$)
such that
$$
\psi =
\alpha_1 e_1 \otimes u_1 + \alpha_2 e_2 \otimes u_2
$$
where $\alpha_i \in {\mathbb C}$
such that
$| \alpha_1|^2 + |\alpha_2|^2=1$.

\subsection{No interference}
Consider the measurement:
\begin{align}
{\mathsf M}_{B({\mathbb C}^2 \otimes H )}
({\mathsf O}_E \otimes {\mathsf O}  , S_{[| \psi \rangle \langle \psi |]} )
\label{eq8}
\end{align}
Then, we see
\begin{itemize}
\item[(F$_1$)]
the probability that a measured value
$(1, x ) (\in \{1\} \times X )$
belongs to
$
\{1\} \times \Xi $
is given by
\end{itemize}
\begin{align*}
&
\langle
\psi,
(I \otimes F(\Xi)  )
\psi
\rangle
\\
=
&
\langle
\alpha_1 e_1 \otimes u_1  + \alpha_2 e_2 \otimes u_2,
(I \otimes F(\Xi)  )
(
\alpha_1 e_1 \otimes u_1  + \alpha_2 e_2 \otimes u_2
)
\rangle
\\
=
&
\langle
\alpha_1 e_1 \otimes u_1  + \alpha_2 e_2 \otimes u_2,
\alpha_1 e_1 \otimes F(\Xi) u_1   + \alpha_2 e_2 \otimes F(\Xi)  u_2
\rangle
\\
=
&
|\alpha_1|^2
\langle
u_1 ,
F(\Xi)u_1
\rangle
+
|\alpha_2|^2
\langle
u_2 ,
F(\Xi)u_2
\rangle
\end{align*}
where the interference term disappears.

\subsection{Interference}
Consider the measurement:
\begin{align}
{\mathsf M}_{B({\mathbb C}^2 \otimes H )}
({\mathsf O}_x \otimes {\mathsf O}  , S_{[| \psi \rangle \langle \psi |]} )
\label{eq9Firs}
\end{align}
Then, we see:
\begin{itemize}
\item[(F$_2$)]
the probability that a measured value
$(1, x ) (\in \{-1,1\} \times X )$
belongs to
$
\{1\} \times \Xi $
is given by
\end{itemize}
\begin{align*}
&
\langle
\psi,
(F_x(\{ 1 \}) \otimes F(\Xi)  )
\psi
\rangle
\\
=
&
\langle
\alpha_1 e_1 \otimes u_1  + \alpha_2 e_2 \otimes u_2,
(F_x(\{ 1 \} \otimes F(\Xi)   ))
(
\alpha_1 e_1 \otimes u_1  + \alpha_2 e_2 \otimes u_2
)
\rangle
\\
=
&
\frac{1}{2}
\langle
\alpha_1 e_1 \otimes u_1  + \alpha_2 e_2 \otimes u_2,
\alpha_1 (e_1+e_2) \otimes F(\Xi) u_1   + \alpha_2 (e_1+ e_2) \otimes F(\Xi)  u_2
\rangle
\\
=
&
\frac{1}{2}
\Big(
|\alpha_1|^2
\langle
u_1 ,
F(\Xi)u_1
\rangle
+
|\alpha_2|^2
\langle
u_2 ,
F(\Xi)u_2
\rangle
+
\overline{\alpha}_1\alpha_2
\langle
u_1 ,
F(\Xi)u_2
\rangle
+
{\alpha}_1 \overline{\alpha}_2
\langle
u_2 ,
F(\Xi)u_1
\rangle
\Big)
\\
=
&
\frac{1}{2}
\Big(
|\alpha_1|^2
\langle
u_1 ,
F(\Xi)u_1
\rangle
+
|\alpha_2|^2
\langle
u_2 ,
F(\Xi)u_2
\rangle
+
2
\mbox{[Real part]}
(
\overline{\alpha}_1\alpha_2
\langle
u_1 ,
F(\Xi)u_2
\rangle
)
\Big)
\end{align*}
where
the
interference term
(i.e.,
the third term)
appears.
\par
Also, we see:
\begin{itemize}
\item[(F$_3$)]
the probability that a measured value
$(-1, x ) (\in \{-1,1\} \times X )$
belongs to
$
\{-1\} \times \Xi $
is given by
\end{itemize}
\begin{align*}
&
\langle
\psi,
(F_x(\{ -1 \}) \otimes F(\Xi)  )
\psi
\rangle
\\
=
&
\langle
\alpha_1 e_1 \otimes u_1  + \alpha_2 e_2 \otimes u_2,
(F_x(\{ -1 \} \otimes F(\Xi)   ))
(
\alpha_1 e_1 \otimes u_1  + \alpha_2 e_2 \otimes u_2
)
\rangle
\\
=
&
\frac{1}{2}
\langle
\alpha_1 e_1 \otimes u_1  + \alpha_2 e_2 \otimes u_2,
\alpha_1 (e_1-e_2) \otimes F(\Xi) u_1   + \alpha_2 (-e_1+ e_2) \otimes F(\Xi)  u_2
\rangle
\\
=
&
\frac{1}{2}
\Big(
|\alpha_1|^2
\langle
u_1 ,
F(\Xi)u_1
\rangle
+
|\alpha_2|^2
\langle
u_2 ,
F(\Xi)u_2
\rangle
-
\overline{\alpha}_1\alpha_2
\langle
u_1 ,
F(\Xi)u_2
\rangle
-
{\alpha}_1 \overline{\alpha}_2
\langle
u_2 ,
F(\Xi)u_1
\rangle
\Big)
\\
=
&
\frac{1}{2}
\Big(
|\alpha_1|^2
\langle
u_1 ,
F(\Xi)u_1
\rangle
+
|\alpha_2|^2
\langle
u_2 ,
F(\Xi)u_2
\rangle
-
2
\mbox{[Real part]}
(
\overline{\alpha}_1\alpha_2
\langle
u_1 ,
F(\Xi)u_2
\rangle
)
\Big)
\end{align*}
where
the
interference term
(i.e.,
the third term)
appears.
\bf
\par
\noindent
Remark 7
\rm
Note that
$$
\underset{\mbox{no interference}}{\fbox{\mbox{(F$_1$)}}}
=
\underset{\mbox{interferences are canceled}}{\fbox{\mbox{(F$_2$)+(F$_3$)}}}
$$
This was experimentally examined in \cite{Walb}.

\section{Wheeler's delayed choice experiment
}
Let $H$ be a two dimensional Hilbert space,
i.e.,
$H={\mathbb C}^2$.
Let $f_1, f_2 \in H$ such that
$$
f_1
=
\bmatrix
1
\\
0
\endbmatrix,
\qquad
f_2
=
\bmatrix
0
\\
1
\endbmatrix
$$
Put
$$
u=\frac{f_1 +f_2}{{\sqrt 2}}
$$
Thus, we have the state
$\rho = |u \rangle \langle u |$
$(\in {\frak S}^p(B({\mathbb C}^2)))$.

Let $U (\in B({\mathbb C}^2 ))$ be
an unitary operator such that
$$
U
=
\bmatrix
1 & 0 \\
0 & e^{i\pi/2 }
\endbmatrix
$$
and let $\Phi: B({\mathbb C}^2) \to B({\mathbb C}^2) $
be the homomorphism such that
$$
\Phi(F) = U^* F U
\qquad
(\forall F \in B({\mathbb C}^2)
)
$$

Consider two observable
${\mathsf O}_f=(\{1,2\}, 2^{\{1,2\}}, F)$
and
${\mathsf O}_g=(\{1,2\}, 2^{\{1,2\}}, G)$
in $B({\mathbb C}^2 )$
such that
\begin{align*}
&
F(\{1\}) = |f_1 \rangle \langle f_1 | ,
\quad F(\{ 2 \}) = |f_2 \rangle \langle f_2 |
\intertext{and}
&
G(\{1\}) = |g_1 \rangle \langle g_1 | ,
\quad G(\{ 2 \}) = |g_2 \rangle \langle g_2 |
\end{align*}
where
\begin{align*}
&
g_1 = \frac{f_1 + f_2 }{\sqrt{2}},
\qquad
g_2 = \frac{f_1 - f_2 }{\sqrt{2}}
\end{align*}

\unitlength=0.7mm
\begin{picture}(150,110)
\put(5,20){{
{
\allinethickness{1.0mm}
{\dottedline{3}(40,80)(50,70)}
{\path(40,30)(50,20)}
{\path(120,70)(110,80)}
\put(115,-10){{\arc{7}{0}{3.14}}$\underset{\;\;\;\; \mbox{(photon counter)}}{\;\;\;\;D_1(=(|f_1 \rangle \langle f_1|))}$}
\put(150,24){{\arc{7}{4.71}{1.57}}$\underset{\;\;\;\; \mbox{(photon counter)}}{\;\;\;\;D_2(=(|f_2 \rangle \langle f_2|))}$}
\allinethickness{0.5mm}
\put(7,74){$\;\;\quad \xrightarrow[]{\frac{1}{\sqrt 2}(f_1+f_2) }\;\;\;\;\;\quad  $}
} 
\put(70,78){$\frac{1}{\sqrt 2}{f_1 }$}
\put(45,75){\vector(0,-1){47}}
\put(48,50){$\frac{{\sqrt{-1}}}{\sqrt 2}{f_2  }$}
\put(115,75){\vector(0,-1){80}}
\put(117,50){$\frac{1}{\sqrt 2}{f_1 }$}
\put(117,5){$\frac{1}{\sqrt 2}{f_1 }$}
\put(45,25){\vector(1,0){100}}
\put(70,28){$\frac{{\sqrt{-1}}}{\sqrt 2}{f_2}$}
\put(130,28){$\frac{{\sqrt{-1}}}{\sqrt 2}{f_2}$}
\put(45,75){\vector(1,0){70}}
}}
\put(40,105){half mirror 1}
\put(40,35){mirror}
\put(120,100){mirror}
\put(35,-5){\color{blue} \bf Figure 3(1). $[D_1 + D_2]$=Observable${\mathsf O}_f$}
\end{picture}
\vskip0.5cm
\par
\noindent
Firstly, consider the measurement:
\begin{align}
{\mathsf M}_{B({\mathbb C}^2)} (\Phi {\mathsf O}_f, S_{[\rho]} ) 
\label{eq10}
\end{align}
Then, we see:
\begin{itemize}
\item[(G$_1$)]
the probability that $\bmatrix \mbox{ a measured value }1 \\ \mbox{ a measured value }2 \endbmatrix$
is obtained by
$
{\mathsf M}_{B({\mathbb C}^2)} (\Phi {\mathsf O}_f, S_{[\rho]} ) 
$
is given by
\begin{align*}
\bmatrix
tr(\rho \cdot \Phi F(\{1\}) )
\\
tr(\rho \cdot \Phi F(\{2\}) )
\endbmatrix
=
\bmatrix
\langle Uu, F(\{1\})Uu \rangle 
\\
\langle Uu, F(\{2\}) Uu \rangle 
\endbmatrix
=
\bmatrix
| \langle Uu, f_1 \rangle|^2 
\\
| \langle Uu, f_2 \rangle|^2
\endbmatrix
=
\bmatrix
\frac{1}{2}
\\
\frac{1}{2}
\endbmatrix
\end{align*}
\end{itemize}
Next, consider the measurement:
\begin{align}
{\mathsf M}_{B({\mathbb C}^2)} (\Phi^2 {\mathsf O}_g, S_{[\rho]} ) 
\label{eq11}
\end{align}
\unitlength=0.7mm
\begin{picture}(150,110)
\put(10,20){{
{
\allinethickness{1.0mm}
{\dottedline{3}(40,80)(50,70)}
{\dottedline{3}(110,30)(120,20)}
{\path(40,30)(50,20)}
{\path(120,70)(110,80)}
\put(115,-10){{\arc{7}{0}{3.14}}$\underset{\;\;\;\; \mbox{(photon counter)}}{\;\;\;\;D_1(=(|g_2 \rangle \langle g_2|))}$}
\put(150,24){{\arc{7}{4.71}{1.57}}$\underset{\;\;\;\; \mbox{(photon counter)}}{\;\;\;\;D_2(=(|g_1 \rangle \langle g_1|))}$}
\allinethickness{0.5mm}
\put(7,74){$\;\;\quad \xrightarrow[]{\frac{1}{\sqrt 2}(f_1+f_2) }\;\;\;\;\;\quad  $}
} 
\put(70,78){$\frac{1}{\sqrt 2}{f_1 }$}
\put(45,75){\vector(0,-1){47}}
\put(48,50){$\frac{{\sqrt{-1}}}{\sqrt 2}{f_2  }$}
\put(115,75){\vector(0,-1){80}}
\put(117,50){$\frac{1}{\sqrt 2}{f_1 }$}
\put(117,5){$\frac{1}{\sqrt 2}{f_1 }-\frac{1}{\sqrt 2}{f_2 }$}
\put(45,25){\vector(1,0){100}}
\put(70,28){$\frac{{\sqrt{-1}}}{\sqrt 2}{f_2}$}
\put(130,28){0}
\put(45,75){\vector(1,0){70}}
}}
\put(40,105){half mirror 1}
\put(90,35){half mirror 2}
\put(40,35){mirror}
\put(120,100){mirror}
\put(35,-5){\color{blue} \bf Figure 3(2). $[D_1 + D_2]$=Observable${\mathsf O}_g$}
\end{picture}
\vskip0.4cm

Then, we see: 
\begin{itemize}
\item[(G$_2$)]
the probability that $\bmatrix \mbox{ a measured value }1 \\ \mbox{ a measured value }2 \endbmatrix$
is obtained by
$
{\mathsf M}_{B({\mathbb C}^2)} (\Phi^2 {\mathsf O}_g, S_{[\rho]} ) 
$
is given by
\begin{align*}
\bmatrix
tr(\rho \cdot \Phi^2 G(\{1\}) )
\\
tr(\rho \cdot \Phi^2 G(\{2\}) )
\endbmatrix
=
\bmatrix
\langle u, \Phi^2 G(\{1\})u \rangle 
\\
\langle u, \Phi^2 G(\{2\}) u \rangle 
\endbmatrix
=&
\bmatrix
| \langle u, UUg_1\rangle |^2
\\
| \langle u, UU g_2 \rangle|^2
\endbmatrix
=
%
\bmatrix
0
\\
1
\endbmatrix
\end{align*}
\end{itemize}

\par
\noindent
Also, consider the following Figure 3(3).
This is clearly the same as the situation of Figure 3(1).
Therefore, this is characterized by the same measurement
$
{\mathsf M}_{B({\mathbb C}^2)} (\Phi {\mathsf O}_f, S_{[\rho]} ) 
$.
\par
\noindent
\unitlength=0.7mm
\begin{picture}(150,110)
\put(10,20){{
{
\allinethickness{1.0mm}
{\dottedline{3}(40,80)(50,70)}
{\dottedline{3}(110,30)(120,20)}
{\path(40,30)(50,20)}
\put(115,-10){{\arc{7}{0}{3.14}}$\underset{\;\;\;\; \mbox{(photon counter)}}{\;\;\;\;D_1(=(|f_2 \rangle \langle f_2|))}$}
\put(120,74){{\arc{7}{4.71}{1.57}}$\underset{\;\;\;\; \mbox{(photon counter)}}{\;\;\;\;D_2(=(|f_1 \rangle \langle f_1|))}$}
\allinethickness{0.5mm}
\put(7,74){$\;\;\quad \xrightarrow[]{\frac{1}{\sqrt 2}(f_1+f_2) }\;\;\;\;\;\quad  $}
} 
\put(70,78){$\frac{1}{\sqrt 2}{f_1 }$}
\put(45,75){\vector(0,-1){47}}
\put(48,50){$\frac{{\sqrt{-1}}}{\sqrt 2}{f_2  }$}
\put(115,18){\vector(0,-1){25}}
\put(117,5){$\frac{-1}{\sqrt 2}{f_2 }$}
\put(45,25){\vector(1,0){68}}
\put(70,28){$\frac{{\sqrt{-1}}}{\sqrt 2}{f_2}$}
\put(45,75){\vector(1,0){70}}
}}
\put(40,105){half mirror 1}
\put(90,35){half mirror 2}
\put(40,35){mirror}
\put(35,-5){\color{blue} \bf Figure 3(3). $[D_2 + D_1]=$Observable${\mathsf O}_f$}
\end{picture}
\vskip0.5cm

%
\par
\noindent
\bf
Remark 8.
\rm
When the half mirror 2 is set in Figure 3(1)(i.e.,
when
the observable ${\mathsf O}_f$ changes to the
${\mathsf O}_g$
),
we see that
the measurement 
$
{\mathsf M}_{B({\mathbb C}^2)} (\Phi {\mathsf O}_f, S_{[\rho]} ) 
$
changes to
$
{\mathsf M}_{B({\mathbb C}^2)} (\Phi^2 {\mathsf O}_g, S_{[\rho]} ) 
$.
Thus, we think that
Wheeler's delayed choice experiment
({\it cf.} \cite{Whee})
is not surprising
in the linguistic interpretation of quantum mechanics.
That is because the problem is not "wave or particle" but
"interference or no interference".
On the other hand, 
the statement ($C_1$) concerning ${\mathsf M}_{B({\mathbb C}^2)} (\Phi {\mathsf O}_f, S_{[\rho]} ) 
$ is surprising, since it implies the non-locality.
This surprising fact is essentially the same as the de Broglie's paradox
(in $B(L^2({\mathbb R}^3))$).

\section{
Hardy's paradox
}

Let $H$ be a two dimensional Hilbert space,
i.e.,
$H={\mathbb C}^2$.
Let $f_1, f_2, g_1, g_2 \in H$ such that
$$
f_1
=
\bmatrix
1
\\
0
\endbmatrix,
\qquad
f_2
=
\bmatrix
0
\\
1
\endbmatrix,
\qquad
g_1 = \frac{f_1 + f_2 }{\sqrt{2}},
\qquad
g_2 = \frac{f_1 - f_2 }{\sqrt{2}}
$$
Put
$$
u=\frac{f_1 +f_2}{{\sqrt 2}} \Big( =g_1 \Big)
$$
Now, consider the tensor Hilbert space
$H \otimes H ={\mathbb C}^2 \otimes {\mathbb C}^2$.
Thus,
put
$$
{\widehat u}=u \otimes u,
\qquad
{\widehat \rho}
=
|u \otimes u \rangle \langle u \otimes u|
$$

\par
\noindent
Define the projection
$P:
{\mathbb C}^2 \otimes {\mathbb C}^2
\to
{\mathbb C}^2 \otimes {\mathbb C}^2
$
such that
$$
P
(
\alpha_{11} f_1 \otimes f_1
+
\alpha_{12} f_1 \otimes f_2
+
\alpha_{21} f_2 \otimes f_1
+
\alpha_{22} f_2 \otimes f_2
)
=
\alpha_{12} f_1 \otimes f_2
+
\alpha_{21} f_2 \otimes f_1
+
\alpha_{22} f_2 \otimes f_2
$$
and thus,
define the
$\Psi: B({\mathbb C}^2 \otimes {\mathbb C}^2) \to
B({\mathbb C}^2 \otimes {\mathbb C}^2)$
by
$$
{\widehat \Psi} ({\widehat A}) = P{\widehat A}P
\qquad
(
{\widehat A} \in B({\mathbb C}^2 \otimes {\mathbb C}^2)
)
$$

\subsection{Concerning the tensor observable ${\mathsf O}_g \otimes {\mathsf O}_g$}
Define the observable $\widehat{\mathsf O}_{gg}=
(\{1,2\} \times \{1,2\}, 2^{\{1,2\} \times \{1,2\}},
\widehat{H}_{gg} )$
in $B({\mathbb C}^2 \otimes {\mathbb C}^2 )$
by
the tensor observable ${\mathsf O}_g \otimes {\mathsf O}_g$,
that is,
\begin{align*}
&
\widehat{H}_{gg}( \{(1,1)\})= |g_1 \otimes g_1 \rangle \langle g_1 \otimes g_1 |,
\quad
\widehat{H}_{gg}( \{(1,2 )\})= |g_1 \otimes g_2 \rangle \langle g_1 \otimes g_2 |,
\\
&
\widehat{H}_{gg}( \{(2,1)\})= |g_2 \otimes g_1 \rangle \langle g_2 \otimes g_1 |,
\quad
\widehat{H}_{gg}( \{(2,2 )\})= |g_2 \otimes g_2 \rangle \langle g_2 \otimes g_2 |
\end{align*}

%
%

Consider the measurement:
\begin{align}
{\mathsf M}_{B({\mathbb C}^2 \otimes {\mathbb C}^2)}
(
{\widehat \Psi}{\widehat{\mathsf O}_{gg}}, S_{[{\widehat \rho}]}
)
\label{eq12}
\end{align}
Then, the probability that a measured value $(2,2)$ is obtained by
$
{\mathsf M}_{B({\mathbb C}^2 \otimes {\mathbb C}^2)}
(
{\widehat \Psi}{\widehat{\mathsf O}}, S_{[{\widehat \rho}]}
)
$
is given by
\begin{align*}
&
\langle  u \otimes u , P{\widehat{H}_{gg}}(\{(2,2)\}) P (u \otimes u) \rangle
\\
=
&
\frac{
|\langle (f_1 - f_2 ) \otimes (f_1 - f_2 )  , f_1 \otimes f_2 + f_2 \otimes f_1 +f_2 \otimes f_2 \rangle|^2 
}{16
}
\\
=
&
\frac{
|\langle 
f_1 \otimes f_1
-
f_1 \otimes f_2 - f_2 \otimes f_1+f_2 \otimes f_2
, 
\;\;
f_1 \otimes f_2 + f_2 \otimes f_1+f_2 \otimes f_2 \rangle|^2 }{
16
}
=
\frac{1}{16}
\end{align*}
Also, the probability that a measured value $(1,1)$ is obtained by
$
{\mathsf M}_{B({\mathbb C}^2 \otimes {\mathbb C}^2)}
(
{\widehat \Psi}{\widehat{\mathsf O}_{gg}}, S_{[{\widehat \rho}]}
)
$
is given by
\begin{align*}
&
\langle  u \otimes u , P{\widehat{H}_{gg}}(\{(1,1)\}) P (u \otimes u) \rangle
\\
=
&
\frac{
|\langle (f_1 + f_2 ) \otimes (f_1 + f_2 )  , f_1 \otimes f_2 + f_2 \otimes f_1 +f_2 \otimes f_2 \rangle|^2 
}{16
}
\\
=
&
\frac{
|\langle 
f_1 \otimes f_1
+
f_1 \otimes f_2 + f_2 \otimes f_1 + f_2 \otimes f_2
, 
\;\;
f_1 \otimes f_2 + f_2 \otimes f_1+f_2 \otimes f_2 \rangle|^2 }{
16
}
=
\frac{9}{16}
\end{align*}
Further, the probability that a measured value $(1,2)$ is obtained by
$
{\mathsf M}_{B({\mathbb C}^2 \otimes {\mathbb C}^2)}
(
{\widehat \Psi}{\widehat{\mathsf O}_{gg}}, S_{[{\widehat \rho}]}
)
$
is given by
\begin{align*}
&
\langle  u \otimes u , P{\widehat{H}_{gg}}(\{(1,2)\}) P (u \otimes u) \rangle
\\
=
&
\frac{
|\langle (f_1 + f_2 ) \otimes (f_1 - f_2 )  , f_1 \otimes f_2 + f_2 \otimes f_1 +f_2 \otimes f_2 \rangle|^2 
}{16
}
\\
=
&
\frac{
|\langle 
f_1 \otimes f_1
-
f_1 \otimes f_2 + f_2 \otimes f_1-f_2 \otimes f_2
, 
\;\;
f_1 \otimes f_2 + f_2 \otimes f_1+f_2 \otimes f_2 \rangle|^2 }{
16
}
=
\frac{1}{16}
\intertext{Similarly,}
&
\langle  u \otimes u , P{\widehat{H}_{gg}}(\{(2,1)\}) P (u \otimes u) \rangle
=
\frac{1}{16}
\end{align*}

\par
\noindent
\bf
Remark 9.
\rm
Recalling Remark 1,
note that
$$
\frac{1}{16}+\frac{9}{16}+\frac{1}{16}+\frac{1}{16}=\frac{3}{4}< 1
$$
Thus,
the probability that no measured value is obtained by the measurement
${\mathsf M}_{B({\mathbb C}^2 \otimes {\mathbb C}^2)}
(
{\widehat \Psi}{\widehat{\mathsf O}}, S_{[{\widehat \rho}]}
) $ is equal to
$\frac{1}{4}$.

\subsection{Concerning the tensor observable ${\mathsf O}_g \otimes {\mathsf O}_f$}
Define the observable $\widehat{\mathsf O}_{gf}=
(\{1,2\} \times \{1,2\}, 2^{\{1,2\} \times \{1,2\}},
\widehat{H}_{gf} )$
in $B({\mathbb C}^2 \otimes {\mathbb C}^2 )$
by
the tensor observable ${\mathsf O}_g \otimes {\mathsf O}_f$,
that is,
\begin{align*}
&
\widehat{H}_{gf}( \{(1,1)\})= |g_1 \otimes f_1 \rangle \langle g_1 \otimes f_1 |,
\quad
\widehat{H}_{gf}( \{(1,2 )\})= |g_1 \otimes f_2 \rangle \langle g_1 \otimes f_2 |,
\\
&
\widehat{H}_{gf}( \{(2,1)\})= |g_2 \otimes f_1 \rangle \langle g_2 \otimes f_1 |,
\quad
\widehat{H}_{gf}( \{(2,2 )\})= |g_2 \otimes f_2 \rangle \langle g_2 \otimes f_2 |
\end{align*}

Consider the measurement:
\begin{align}
{\mathsf M}_{B({\mathbb C}^2 \otimes {\mathbb C}^2)}
(
{\widehat \Psi}{\widehat{\mathsf O}_{gf}}, S_{[{\widehat \rho}]}
)
\label{eq13}
\end{align}
Then, the probability that a measured value $(2,2)$ is obtained by
$
{\mathsf M}_{B({\mathbb C}^2 \otimes {\mathbb C}^2)}
(
{\widehat \Psi}{\widehat{\mathsf O}}_{gf}, S_{[{\widehat \rho}]}
)
$
is given by
\begin{align*}
&
\langle  u \otimes u , P{\widehat{H}_{gf}}(\{(2,2)\}) P (u \otimes u) \rangle
\\
=
&
\frac{
|\langle (f_1 - f_2 ) \otimes f_2   , f_1 \otimes f_2 + f_2 \otimes f_1 +f_2 \otimes f_2 \rangle|^2 
}{8
}
=0
\end{align*}
Also, the probability that a measured value $(1,1)$ is obtained by
$
{\mathsf M}_{B({\mathbb C}^2 \otimes {\mathbb C}^2)}
(
{\widehat \Psi}{\widehat{\mathsf O}_{gf}}, S_{[{\widehat \rho}]}
)
$
is given by
\begin{align*}
&
\langle  u \otimes u , P{\widehat{H}_{gf}}(\{(1,1)\} )P (u \otimes u) \rangle
\\
=
&
\frac{
|\langle (f_1 + f_2 ) \otimes f_1  , f_1 \otimes f_2 + f_2 \otimes f_1 +f_2 \otimes f_2 \rangle|^2 
}{8
}
=
\frac{1}{8}
\end{align*}
Further, the probability that a measured value $(1,2)$ is obtained by
$
{\mathsf M}_{B({\mathbb C}^2 \otimes {\mathbb C}^2)}
(
{\widehat \Psi}{\widehat{\mathsf O}_{gf}}, S_{[{\widehat \rho}]}
)
$
is given by
\begin{align*}
&
\langle  u \otimes u , P{\widehat{H}_{gf}}(\{(1,2)\}) P (u \otimes u) \rangle
\\
=
&
\frac{
|\langle (f_1 + f_2 ) \otimes f_2 , f_1 \otimes f_2 + f_2 \otimes f_1 +f_2 \otimes f_2 \rangle|^2 
}{16
}
=
\frac{4}{8}
\intertext{Similarly,}
&
\langle  u \otimes u , P{\widehat{H}_{gf}}(\{(2,1)\}) P (u \otimes u) \rangle
\\
=
&
\frac{
|\langle (f_1 - f_2 ) \otimes f_1 , f_1 \otimes f_2 + f_2 \otimes f_1 +f_2 \otimes f_2 \rangle|^2 
}{8
}
=
\frac{1}{8}
\end{align*}

\par
\noindent
\bf
Remark 10.
\rm
It is usual to consider that "Which way pass problem"
is nonsense.
However,
for the other aspect of this problem,
see Remarks 11 and 12 later.

\section{
The three boxes paradox
}

Let $H$ be the three dimensional Hilbert space,
i.e.,
$H={\mathbb C}^3$.
Let $f_1, f_2, f_3 \in H$ such that
$$
f_1
=
\bmatrix
1
\\
0
\\
0
\endbmatrix,
\qquad
f_2
=
\bmatrix
0
\\
1
\\
0
\endbmatrix,
\qquad
f_3
=
\bmatrix
0
\\
0
\\
1
\endbmatrix
$$
Put
$$
u=\frac{f_1 +f_2 + f_3}{{\sqrt 3}},
\qquad
g_1=
\frac{f_1 +f_2 - f_3}{{\sqrt 3}}.
$$
And,
put
$$
{\rho}
=
|u \rangle \langle u |
$$
Further, consider two observables
${\mathsf O}_1=( \{1,2 \}, 2^{\{1,2 \}}, G)$
and
${\mathsf O}_2=( \{1,2,3 \}, 2^{\{1,2,3 \}}, F)$
in $B(H)$
such that
$$
G(\{1 \})= |g_1 \rangle \langle g_1 |
=
\frac{1}{3}
|f_1 +f_2 - f_3 \rangle \langle f_1 +f_2 - f_3 | 
(\equiv P_1 ),
\quad
G(\{2 \})= I- |g_1 \rangle \langle g_1 |
(\equiv P_2 )
$$
and
$$
F(\{1 \})= |f_1 \rangle \langle f_1 |,
\quad
F(\{3 \})= |f_2 \rangle \langle f_2 |,
\quad
F(\{3 \})= |f_3 \rangle \langle f_3 |,
$$
And consider the measurements
\begin{align}
{\mathsf M}_{B(H)}( {\mathsf O}_1, S_{[\rho]} )
\;\;\;\mbox{ and }\;\;\;
{\mathsf M}_{B(H)}( {\mathsf O}_2, S_{[\rho]} )
\label{eq14}
\end{align}
Clearly, the probability that a measured value $1$ obtained by
${\mathsf M}_{B(H)}( {\mathsf O}_1, S_{[\rho]} )$
is given by
$$
\langle u,G(\{1\}) u \rangle
=
\langle u, |g_1 \rangle \langle g_1 | u \rangle
=
\frac{1}{9} |\langle f_1 +f_2 + f_3, f_1 +f_2 - f_3 \rangle |^2
=
\frac{1}{9}
$$
and,
the probability that a 
$\bmatrix 
\mbox{measured value }1 \\
\mbox{measured value }2 \\
\mbox{measured value }3
\endbmatrix
$
obtained by
${\mathsf M}_{B(H)}( {\mathsf O}_2, S_{[\rho]} )$
is given by
$$
\bmatrix
\langle u, |f_1 \rangle \langle f_1 | u \rangle
=
 |\langle f_1, u \rangle |^2
=
\frac{1}{3}
\\
\langle u, |f_2 \rangle \langle f_2 | u \rangle
=
|\langle f_2, u \rangle |^2
=
\frac{1}{3}
\\
\langle u, |f_3 \rangle \langle f_3 | u \rangle
=
|\langle f_3, u \rangle |^2
=
\frac{1}{3}
\endbmatrix
$$
Since
${\mathsf O}_1$
and
${\mathsf O}_2$
do not commute,
the simultaneous observable
${\mathsf O}_1 \times {\mathsf O}_2$
does not exist.
However, putting
$X=\{1,2 \}$,
$Y=\{1,2,3 \}$,
${\mathsf O}_1 \times {\mathsf O}_2$
may be formally written by
$$
{\mathsf O}_1 \times {\mathsf O}_2
=
(X \times Y, 2^{X \times Y}, G \times  F  )
$$
where
\begin{align*}
&
G(\{1\})F(\{1\})  =(\langle g_1, f_1 \rangle) |g_1 \rangle \langle f_1 |
=
\frac{1}{3} |f_1 + f_2 - f_3  \rangle \langle f_1 |
\\
&
G(\{1\})F(\{2\}) =(\langle g_1, f_2 \rangle) |g_1 \rangle \langle f_2 |
=
\frac{1}{3} |f_1 + f_2 - f_3 \rangle \langle f_2 |
\\
&
G(\{1\})F(\{3\}) =(\langle g_1, f_3 \rangle) |g_1 \rangle \langle f_3 |
=
\frac{1}{3} |f_1 + f_2 - f_3 \rangle \langle f_3 |
\\
&
\qquad \cdots \cdots
\end{align*}
However, we try to consider the "measurement"
\begin{align}
"{\mathsf M}_{B(H)}({\mathsf O}_1 \times{\mathsf O}_2 , S_{[\rho]} )".
\label{eq15}
\end{align}
And further, we can calculate as follows.
\begin{itemize}
\item[(H)]
under the condition that the measured value $(1, y)$ is obtained by
"${\mathsf M}_{B(H)}({\mathsf O}_1 \times{\mathsf O}_2 \times {\mathsf O}_2, S_{[\rho]} )$",
the probability that $\bmatrix
y=1
\\
y=2
\\
y=3
\endbmatrix
$
is
formally
given by
$$
\bmatrix
\frac{\langle u , G(\{ 1\}) F(\{1\} ) u \rangle}{\langle u , G(\{ 1\})u \rangle}=
\frac{\langle u , (\langle g_1, f_1 \rangle) |g_1 \rangle \langle f_1 | u 
\rangle}{\langle u, |g_1 \rangle \langle g_1 | u \rangle}
=
\frac{\langle g_1, |f_1 \rangle \langle f_1 | u 
\rangle}{\langle g_1 | u \rangle}
=1
\\
\frac{\langle u , G(\{ 1\}) F(\{2\} ) u \rangle}{\langle u , G(\{ 1 \})u \rangle}=
\frac{\langle u , (\langle g_1, f_2 \rangle) |g_1 \rangle \langle f_2 | u 
\rangle}{\langle u, |g_1 \rangle \langle g_1 | u \rangle}
=
\frac{\langle g_1, |f_2 \rangle \langle f_2 | u 
\rangle}{\langle g_1 | u \rangle}
=1
\\
\frac{\langle u , G(\{ 1\}) F(\{3\} ) u \rangle}{\langle u , G(\{ 1\})u \rangle}=
\frac{\langle u , (\langle g_1, f_3 \rangle) |g_1 \rangle \langle f_3 | u 
\rangle}{\langle u, |g_1 \rangle \langle g_1 | u \rangle}
=
\frac{\langle g_1, |f_3 \rangle \langle f_3 | u 
\rangle}{\langle g_1 | u \rangle}
=-1
\endbmatrix
$$
\end{itemize}
which shows the strange fact
(i.e.,
"minus probability").

\par
\noindent
\bf
Remark 11.
\rm
Since
${\mathsf O}_1$
and
${\mathsf O}_2$
do not commute,
${\mathsf O}_1 \times {\mathsf O}_2$
is not an observable, but $B(H)$-valued measure space
({\it cf.}
Remark 5).
Thus, it is usual to consider that the above (H) is meaningless.
However,
if some will find the idea such that the (H) becomes meaningful,
then
the idea should be added to the linguistic interpretation
mentioned in Section 1.4.
For example,
the idea in ref. \cite{Ahar}
(the weak value associated with a weak measurement)
may be somewhat hopeful,
but we can not assure it.

\par
\noindent
\bf
Remark 12
\rm
(Continued from Hardy's paradox in Section 5).
Define the observable $\widehat{\mathsf O}_{ff}=
(\{1,2\} \times \{1,2\}, 2^{\{1,2\} \times \{1,2\}},
\widehat{H}_{ff} )$
in $B({\mathbb C}^2 \otimes {\mathbb C}^2 )$
by
the tensor observable ${\mathsf O}_f \otimes {\mathsf O}_f$,
that is,
\begin{align*}
&
\widehat{H}_{ff}( \{(1,1)\})= |f_1 \otimes f_1 \rangle \langle f_1 \otimes f_1 |,
\quad
\widehat{H}_{ff}( \{(1,2 )\})= |f_1 \otimes f_2 \rangle \langle f_1 \otimes f_2 |,
\\
&
\widehat{H}_{ff}( \{(2,1)\})= |f_2 \otimes f_1 \rangle \langle f_2 \otimes f_1 |,
\quad
\widehat{H}_{ff}( \{(2,2 )\})= |f_2 \otimes f_2 \rangle \langle f_2 \otimes f_2 |
\end{align*}
In spite of the non-commutativity of
$
{\widehat \Psi}{\widehat{\mathsf O}_{gg}}$
and
$
\widehat{\mathsf O}_{ff}
$,
consider the "measurement":
\begin{align}
{\mathsf M}_{B({\mathbb C}^2 \otimes {\mathbb C}^2)}
(
{\widehat \Psi}{\widehat{\mathsf O}_{gg}} \times \widehat{\mathsf O}_{ff} , S_{[{\widehat \rho}]}
)
\label{eq16}
\end{align}
And we can calculate as follows.
\begin{itemize}
\item[(I)]
under the condition that the measured value $((2,2),(y_1,y_2))$ is obtained by
"${\mathsf M}_{B({\mathbb C}^2 \otimes {\mathbb C}^2)}
(
{\widehat \Psi}{\widehat{\mathsf O}_{gg}} \times \widehat{\mathsf O}_{ff} , S_{[{\widehat \rho}]}
)$",
the probability (or precisely, weak value) that $\bmatrix
(y_1,y_2)=(1,1)
\\
(y_1,y_2)=(1,2)
\\
(y_1,y_2)=(2,1)
\\
(y_1,y_2)=(2,2)
\endbmatrix
$
is
formally
given by
\end{itemize}
\begin{align*}
&
\bmatrix
\frac{\langle u \otimes u , P {\widehat H}_{gg}  (\{ (2,2) \})P {\widehat H}_{ff} (\{(1,1) \} ) (u \otimes u) \rangle}{
\langle u \otimes u , P {\widehat H}_{gg}  (\{ (2,2) \})P
(u \otimes u) \rangle }
\\
\frac{\langle u \otimes u , P {\widehat H}_{gg}  (\{ (2,2) \})P {\widehat H}_{ff} (\{(1,2) \} ) (u \otimes u) \rangle}{
\langle u \otimes u , P {\widehat H}_{gg}  (\{ (2,2) \})P
(u \otimes u) \rangle }
\\
\frac{\langle u \otimes u , P {\widehat H}_{gg}  (\{ (2,2) \})P {\widehat H}_{ff} (\{(2,1) \} ) (u \otimes u) \rangle}{
\langle u \otimes u , P {\widehat H}_{gg}  (\{ (2,2) \})P
(u \otimes u) \rangle }
\\
\frac{\langle u \otimes u , P {\widehat H}_{gg}  (\{ (2,2) \})P {\widehat H}_{ff} (\{(2,2) \} ) (u \otimes u) \rangle}{
\langle u \otimes u , P {\widehat H}_{gg}  (\{ (2,2) \})P
(u \otimes u) \rangle }
\endbmatrix
\\
=
&
{4}
{
\small
\bmatrix
\langle (f_1+f_2) \otimes (f_1+f_2) , P|g_2 \otimes g_2 \rangle \langle g_2 \otimes g_2 |P |
f_1 \otimes f_1 \rangle
\langle
f_1 \otimes f_1, (f_1+f_2) \otimes (f_1+f_2) \rangle
\\
\langle (f_1+f_2) \otimes (f_1+f_2) , P|g_2 \otimes g_2 \rangle \langle g_2 \otimes g_2 |P |
f_1 \otimes f_2 \rangle
\langle
f_1 \otimes f_2, (f_1+f_2) \otimes (f_1+f_2) \rangle
\\
\langle (f_1+f_2) \otimes (f_1+f_2) , P|g_2 \otimes g_2 \rangle \langle g_2 \otimes g_2 |P |
f_2 \otimes f_1 \rangle
\langle
f_2 \otimes f_1, (f_1+f_2) \otimes (f_1+f_2) \rangle
\\
\langle (f_1+f_2) \otimes (f_1+f_2) , P|g_2 \otimes g_2 \rangle \langle g_2 \otimes g_2 |P |
f_2 \otimes f_2 \rangle
\langle
f_2 \otimes f_2, (f_1+f_2) \otimes (f_1+f_2) \rangle
\endbmatrix
}
\\
=
&
{
\small
\bmatrix
\langle (f_1+f_2) \otimes (f_1+f_2) , -f_1 \otimes f_2-f_2 \otimes f_1 + f_2 \otimes f_2 \rangle \langle -f_1 \otimes f_2-f_2 \otimes f_1 + f_2 \otimes f_2,
f_1 \otimes f_1 \rangle
\\
\langle (f_1+f_2) \otimes (f_1+f_2) , -f_1 \otimes f_2-f_2 \otimes f_1 + f_2 \otimes f_2 \rangle \langle -f_1 \otimes f_2-f_2 \otimes f_1 + f_2 \otimes f_2,
f_1 \otimes f_2 \rangle
\\
\langle (f_1+f_2) \otimes (f_1+f_2) , -f_1 \otimes f_2-f_2 \otimes f_1 + f_2 \otimes f_2 \rangle \langle -f_1 \otimes f_2-f_2 \otimes f_1 + f_2 \otimes f_2,
f_2 \otimes f_1 \rangle
\\
\langle (f_1+f_2) \otimes (f_1+f_2) , -f_1 \otimes f_2-f_2 \otimes f_1 + f_2 \otimes f_2 \rangle \langle -f_1 \otimes f_2-f_2 \otimes f_1 + f_2 \otimes f_2,
f_2 \otimes f_2 \rangle
\endbmatrix
}
\\
=
&
{
\small
\bmatrix
0
\\
(-1) \times (-1)
\\
(-1) \times (-1)
\\
(-1) \times (1)
\endbmatrix
=
\bmatrix
0
\\
1
\\
1
\\
-1
\endbmatrix
}
\end{align*}
This (I) and the idea in ref.\cite{Ahar} are superficially similar,
but
completely different in essence.
However, if the latter says something good,
we can expect that the (I) is somewhat meaningful.  
For completeness, note that
quantum language is not physics but language.
Therefore,
we say that
\begin{itemize}
\item[(J)]
if this statement (I) can be used effectively, then the concept: "weak value" should be accepted in the linguistic interpretation,
however,
if this is not more than "even not wrong",
we will not be concerned with "the weak value".
\end{itemize}


\section{Conclusions}
In this paper,
we discussed
the double slits experiment,
the quantum eraser experiment,
Wheeler's delayed choice experiment,
Hardy's paradox
and the three boxes paradox
in the linguistic interpretation of quantum mechanics.

%

Quantum language says that
everything should be described in terms of
Axioms 1 and 2.
Therefore,
we always have to describe "measurement" explicitly.
In fact, in this paper, any measurement
was explicitly described such as
the formula
[(\ref{eq5})-(\ref{eq14}),
(\ref{eq15}),
(\ref{eq16})
].
Particularly,
in Section 2,
we say that 
the double-slit experiment
can not be understood without the concept of
"branch".
And, in Section 4,
we note that
Wheeler's delayed choice experiment
is not surprising,
since it
should be regarded as the problem
such as
"interference or no interference".

Through these arguments, we assert that the linguistic interpretation is just 
the final version of so called Copenhagen interpretation.
And therefore, the Copenhagen interpretation does not belong to physics
but the linguistic world view ({\it cf.} \hyperlink{Figure 1}{Figure 1}).

\par
We hope that our proposal will be discussed and examined from various view-points.

\rm
\par


\rm
\par
\renewcommand{\refname}{
\large 
References}
{
\small

\normalsize
}

\end{document}